\documentclass[reprint,showkeys,showpacs,preprintnumbers,nofootinbib,amsmath,amssymb,aps]{revtex4-1}

\usepackage{graphicx}
\usepackage{dcolumn}
\usepackage{bm}

\def\nq{\hspace*{-1em}}

\def\cm{\hspace*{1cm}}

\def\Jl#1#2{#1 {\bf #2},\ }

\def\ApJ#1 {\Jl{Astroph. J.}{#1}}
\def\CQG#1 {\Jl{Class. Quantum Grav.}{#1}}
\def\DAN#1 {\Jl{Dokl. AN SSSR}{#1}}
\def\GC#1 {\Jl{Grav. Cosmol.}{#1}}
\def\GRG#1 {\Jl{Gen. Rel. Grav.}{#1}}
\def\JETF#1 {\Jl{Zh. Eksp. Teor. Fiz.}{#1}}
\def\JETP#1 {\Jl{Sov. Phys. JETP}{#1}}
\def\JHEP#1 {\Jl{JHEP}{#1}}
\def\JMP#1 {\Jl{J. Math. Phys.}{#1}}
\def\NPB#1 {\Jl{Nucl. Phys. B}{#1}}
\def\NP#1 {\Jl{Nucl. Phys.}{#1}}
\def\PLA#1 {\Jl{Phys. Lett. A}{#1}}
\def\PLB#1 {\Jl{Phys. Lett. B}{#1}}
\def\PRD#1 {\Jl{Phys. Rev. D}{#1}}
\def\PRL#1 {\Jl{Phys. Rev. Lett.}{#1}}

\def\lal{&&{}}
\def\eq{Eq.\,}
\def\eqs{Eqs.\,}
\def\beq{\begin{equation}}
\def\eeq{\end{equation}}
\def\bear{\begin{eqnarray}}
\def\bearr{\begin{eqnarray} \lal}
\def\ear{\end{eqnarray}}
\def\earn{\nonumber \end{eqnarray}}

\def\yyy{\\[5pt] \lal }

\def\dst{\displaystyle}

\def\fracd#1#2{{\dst\frac{#1}{#2}}}

\def\Half{{\fracd{1}{2}}}

\def\const{{\rm const}}

\def\eqn#1{\eq\eqref{#1}}
\def\rf{\eqref}
\def\mn{_{\mu\nu}}
\def\MN{^{\mu\nu}}

\def\kappa{\varkappa}

\def\ssph{static, spherically symmetric}

\begin{document}

\title{Comment on ``Construction of regular black holes in general relativity''} 

\author{Kirill A. Bronnikov}
\affiliation{VNIIMS, Ozyornaya ul. 46, Moscow 119361, Russia}
\affiliation{Institute of Gravitation and Cosmology, Peoples' Friendship University of Russia 
   (RUDN University), 
              ul. Miklukho-Maklaya 6, Moscow 117198, Russia}
\affiliation{National Research Nuclear University ``MEPhI'', Kashirskoe sh. 31, Moscow 115409, Russia}
\email{kb20@yandex.ru}

\date{}

\begin{abstract}
   {It is claimed that the paper by Zhong-Ying Fan and Xiaobao Wang 
     [\PRD {94} 124027 (2016); arXiv: 1610.02636]
    on nonlinear electrodynamics coupled to general relativity, being correct in general, in some
    respects repeats previously obtained results without giving proper references. 
    There is also an important point missing in this paper, but necessary for understanding the 
    physics of the system: in solutions with an electric charge, a regular center requires a 
    non-Maxwell behavior of Lagrangian function $L(f),\ (f= F\mn F\MN)$ at small $f$. 
    Therefore, in all electric regular black hole solutions with a Reissner-Nordstr\"om asymptotic, 
    the Lagrangian $L(f)$ is different in different parts of space, and the electromagnetic field
    behaves in a singular way at surfaces where $L(f)$ suffers branching. 
   }

\end{abstract}


\maketitle

  Nonlinear electrodynamics (NED) as a possible material source in general relativity (GR) and 
  its extensions attracts much attention since, among other reasons, it leads to many space-time
  geometries of interest, in particular, regular black holes and starlike, or solitonlike configurations.

  The paper by Fan and Wang \cite{FW} belongs to this trend but in some important points repeats already 
  known results, and many relevant papers are absent in the list of references. There are also 
  some well-known important physical properties of solutions with an electric charge, which are not 
  mentioned in \cite{FW} but deserve mentioning as necessary information for readers (e.g.,
  students) who are not experts in the field.  
  
  To begin with, the key inferences of \cite{FW} are based on the general \ssph\ solution of GR 
  coupled to NED in the case of an electric field obtained in 1969 by Pellicer and Torrence \cite{Pel-T},
  not cited in \cite{FW}. This consideration was extended in \cite{k-NED} to systems containing 
  both electric and magnetic charges. For a further discussion, let us briefly reproduce it here.

  In GR coupled to NED one considers the action
\beq            \label{S}
	S = \Half \int \sqrt {-g} d^4 x [R - L (f)], 	\cm   f = F\mn F\MN 
\eeq
  ($F\mn$ is the Maxwell tensor, units with $c = 8\pi G =1$ are used) with an arbitrary function 
  $L (f)$. Then, assuming static spherical symmetry, the stress-energy tensor (SET) satisfies 
  the condition $T^t_t = T^r_r$, hence, due to the Einstein equations, the metric can be written as 
\beq            \label{ds} 
	ds^2 = A(r) dt^2 - dr^2/A(r) - r^2 (d\theta^2+\sin^2 \theta d\phi^2).
\eeq
  The only nonzero components of $F\mn$ are $F_{tr} =- F_{rt}$ (a radial electric field) 
  and $F_{\theta\phi} = - F_{\phi\theta}$ (a radial magnetic field). The Maxwell-like equations
  $\nabla_\mu (L_f F\MN) = 0$ and the Bianchi identities $\nabla_\mu {}^*F\MN = 0$ give
\beq              \label{F_mn}
                      r^2 L_f F^{tr} = q_e, \cm    F_{\theta\phi} = q_m\sin\theta,
\eeq
  where $q_e$ and $q_m$ are the electric and magnetic charges, respectively, and $L_f \equiv dL/df$. 
  Accordingly, the nonzero SET components are 
\bearr  \nq         \label{SET}
            T^t_t = T^r_r = \Half L + f_e L_f, \quad T^\theta_\theta = T^\phi_\phi = \Half  L - f_m L_f,
\yyy    \nq
	f_e = 2 F_{tr}F^{rt}=\frac{2q_e^2}{L_f^2 r^4}, \ \ \ 
	f_m = 2 F_{\theta\phi}F^{\theta\phi} = \frac {2q_m^2}{r^4},             \label{f_em}
\ear
  so that the invariant $f$ is $f = f_m - f_e$. The metric function $A(r)$ is found from the 
  Einstein equations as 
\beq          \label{A}
	A(r) = 1 -\frac{2M(r)}{r}, \qquad M(r) = \frac 12 \int T^t_t(r) r^2 dr,  
\eeq
  where $T^t_t$ is the energy density. It is a relation including both electric
  and magnetic fields, written in a general form \cite{k-NED}.

  Fan and Wang \cite{FW} claim that they have presented a general procedure for constructing 
  exact regular black hole solutions with electric or magnetic charges in GR coupled to NED.
  However, this procedure was in fact already described in \cite{k-NED}.
  Indeed, for a magnetic solution ($q_e=0$), given any $A(r)$, using \eqn{A}, one easily 
  calculates $T^t_t = L/2$ as a function of $r$, and $f = f_m(r)$ is known from \rf{f_em}, 
  thus $L(f)$ is also determined. On the contrary, starting from $L(f)$ and using \rf{f_em}, we
  obtain $M(r)$ and $A(r)$ from \rf{A}. Moreover, a necessary condition for obtaining a regular 
  center is that $L(f)$ should tend to a finite limit as $f\to\infty$ \cite{k-NED} (an observation 
  absent in \cite{FW}). On the other hand, the description in \cite{FW} contains some 
  additional relations, details and explanations.

  Electric solutions are obtained in a similar way using the ``Hamiltonian'' formulation of NED 
  (see, e.g., \cite{AB-G}), produced from the original one by a Legendre transformation: one 
  introduces the tensor $P\mn = L_f F\mn$ with its invariant $p = -P\mn P\MN$ and considers 
  the Hamiltonian-like quantity $H = 2f L_f  - L = 2T^t_t$ as a function of $p$; then $H(p)$ can 
  be used to specify the whole theory. One has then
\beq
          L = 2p H_p - H, \quad  L_f H_p = 1, \quad  f = p H_p^2.
\eeq
  with $H_p \equiv dH/dp$. Then for ``electric'' solutions ($q=q_e\ne 0$, $q_m=0$), specifying 
  $H(p)$, we directly find $M(r)$ and $A(r)$ using \rf{A} since we have simply $p= 2q^2/r^4$. 
  If we specify $A(r)$, from \rf{A} we easily find $H(p)$. All this was described in \cite{k-NED},
  see there \eqs (12) for magnetic solutions and (19) for electric ones.

  In both cases, selection of special families of such solutions governed by a few parameters 
  (as is done in \cite{FW}) is quite an easy task since the function $A(r)$, specifying the solutions,
  is arbitrary, hence the number of free parameters can also be arbitrary. One should only take care
  of the boundary condition $A(r) = 1 + O(r^2)$ as $r \to 0$ if a regular center is required, and 
  provide $A(r) \approx 1 - 2M/r$ as  $r\to\infty$ ($M=\const$) to have a Schwarzschild asymptotic.
  If this $A(r)$ has zeros, corresponding to horizons, it is a BH solution, while if everywhere $A > 0$,
  it is a particlelike or soliton solution (the latter opportunity is not mentioned in \cite{FW}). 

  An important point concerning electric solutions is the existence of a no-go theorem \cite{B-Shi} 
  (which was probably unknown to the authors of \cite{FW}), 
  saying that {\it there is no such Lagrangian function $L(f)$ having a Maxwell
  weak-field limit ($L\sim f$ as $f\to 0$) that the electric solution \rf{ds}, \rf{F_mn}, \rf{A} has a 
  regular center.} The reason is that at such a center the electric field should be zero but the field 
  equations then imply $f L_f^2 \to \infty$, hence $L_f \to \infty$ as $r\to 0$.
  It was further shown in \cite{k-NED} that a regular center is also impossible in dyonic 
  configurations, with both $q_e\ne 0$ and $q_m \ne 0$, if $L(f)$ has a Maxwell weak-field limit.

  An alternative (but equivalent) formulation of this no-go theorem is that {\it if a \ssph\ solution to the 
  theory \rf{S} with $q_e \ne 0$ contains a regular center, then $L(f)$ is non-Maxwell at small $f$.} 

  A natural question is: how does this no-go theorem combine with the existing examples
  of regular electric solutions, e.g., the one given in \cite{AB-G} and others, mentioned or cited
  in \cite{FW}? An answer was given in \cite{B-comment}: in all such cases, in a ``regular'' 
  solution there are different Lagrangian functions $L(F)$ at large and small $r$. 
  At large $r$, where $f\to 0$, we have $L\sim f$ whereas at small $r$ the theory is strongly 
  non-Maxwell ($f\to 0$ but $L_f \to \infty$), in agreement with the no-go theorem. 
  An inspection showed that it is indeed the case in all examples. 

  According to \cite{k-NED, B-comment}, in the ``Hamiltonian'' framework,  
  at a regular center we have $p\to \infty$ but a finite limit of $H$, and the integral in \rf{A}  
  gives the mass function and $A(r)$. However, in all regular solutions where $f = 0$ at both $r=0$ 
  and $r=\infty$, the function $f$ inevitably has at least one maximum at some $p=p^*$, violating the 
  monotonicity of $f(p)$, which is necessary for equivalence of the $f$ and $p$ frameworks. 
  It has been shown \cite{k-NED} that at an extremum of $f(p)$ the Lagrangian function $L(f)$ 
  suffers branching, its plot forming a cusp, and different functions $L(f)$ correspond to  
  $p < p^*$ and $p > p^*$. Another kind of branching occurs at extrema of $H(p)$, if any, and 
  the number of Lagrangians $L(f)$ on the way from infinity to the center equals the number 
  of  monotonicity ranges of $f(p)$. 

  It was mentioned in \cite{FW} that ``the original $L(f)$ formalism may not be
  appropriate any longer in this case because one will end with a multivalued $L(f)$, which has 
  different branches for a well-defined single one $H(p)$.'' It should be stressed, however, 
  that this branching is an {\sl inevitable\/} property of all regular electric solutions with a
  Reissner-Nordstr\"om asymptotic behavior. 

  It might seem that the ``Hamiltonian'' framework is not worse than the Lagrangian one, even
  though the latter is directly related to the least action principle. However, as shown in 
  \cite{k-NED}, at $p=p^*$ the electromagnetic field exhibits a singular behavior, well revealed 
  using the effective metric \cite{Nov-1, Nov-2} in which NED photons move along null geodesics.
  This metric is singular at extrema of $f(p)$, and the effective potential for geodesics exhibits 
  infinitely deep wells where NED photons are infinitely blueshifted \cite{k-NED, Nov-2} and can 
  after all create a curvature singularity due to back-reaction on the metric. Thus 
  any such solution not only fails to correspond to a fixed Lagrangian $L(f)$ but has 
  other important undesired features. In my opinion, it is a necessary addition to the description
  of electric solutions in \cite{FW}.  
  
  Is it possible to circumvent the above no-go theorem for electric solutions? The answer is yes
  \cite{Bur1}: one can consider a kind of phase transition on a certain sphere, outside which 
  there is a purely electric field $F\mn$ but inside which the field is purely magnetic. An external observer 
  then sees an electrically charged BH or soliton. 

  Fan and Wang \cite{FW} also describe a straightforward extension of \ssph\ NED solutions to 
  GR with a nonzero negative cosmological constant $\Lambda$, leading to their anti-de Sitter 
  asymptotic behavior; however, this extension (with both positive and negative $\Lambda$)  has 
  been already considered,  e.g., in \cite{Mat-09, Mat-13, Fer-15}. 
  Actually, if we add $-2\Lambda$ to $R$ in the action 
  \rf{S}, the only change in the expression \rf{A} for the metric is that the term  
  $ - \Lambda r^2/3$ is added to $A(r)$. With or without $\Lambda$, if $A(r)$ is known 
  (or chosen by hand), the form of the theory is easily restored from \rf{A}: $dM(r)/dr$ directly 
  gives $H(p)$ for electric configurations or $L(f)$ for magnetic ones since $p(r)$ or $f(r)$, 
  respectively, are known in these cases. On the contrary, knowing $L(f)$ or $H(p)$, it is easy to 
  find $A(r)$ in magnetic or electric configurations, respectively. 

  To summarize, there is a substantial gap in \cite{FW}, connected with the fact that the  
  ``regular black hole construction procedure'' was already described earlier. An important point
  missing in \cite{FW} is the inevitable undesired property of regular electric solutions if one 
  requires a Maxwell weak field limit of NED at large radii (multivaluedness of the Lagrangian
  function $L(f)$ and troubles with the electromagnetic field at its branching points). Somewhat 
  less important is a missing mentioning of possible solitonic and asymptotically de Sitter solutions. 
  An evident shortcoming is the absence of necessary references,  directly related to the subject, 
  such as \cite{Pel-T, k-NED, Bur1, Mat-09, Mat-13, Fer-15}, and maybe some others. 

  Does all that mean that there are no new results of interest in \cite{FW}? Certainly not. 
  New examples of regular BH solutions both with and without a cosmological constant are 
  obtained and discussed, some useful general relations have been obtained for the GR-NED 
  set of equations, and the whole section V entitled ``The first law of thermodynamics'' is quite 
  interesting and is not restricted to the first law only: there are a generalization of Smarr's
  formula and new expressions for the entropy products. So, despite the above criticism, this paper 
  seems to be quite a useful contribution to the studies of regular black holes.
   
\subsection*{Acknowledgments}

  I thank Milena Skvortsova and Sergei Bolokhov for helpful discussions. 
  The work was partly performed within the framework of the Center FRPP 
  supported by MEPhI Academic Excellence Project (contract No. 02.a03.21.0005, 27.08.2013).
  The work was also funded by the RUDN University Program 5-100.


\end{document}